\documentclass[sigconf]{acmart}

\AtBeginDocument{%
  }

\copyrightyear{2022}
\acmYear{2022}
\setcopyright{licensedothergov}\acmConference[ACSAC '22]{Annual Computer Security Applications Conference}{December 5--9, 2022}{Austin, TX, USA}
\acmBooktitle{Annual Computer Security Applications Conference (ACSAC '22), December 5--9, 2022, Austin, TX, USA}
\acmPrice{15.00}
\acmDOI{10.1145/3564625.3564659}
\acmISBN{978-1-4503-9759-9/22/12}
 
\acmBadgeR[https://www.acm.org/publications/policies/artifact-review-and-badging-current]{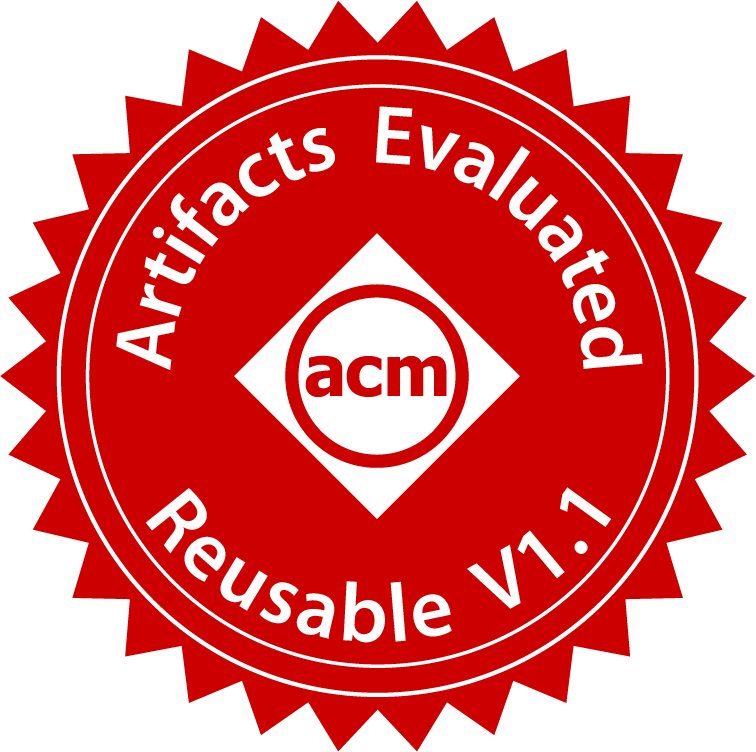}

\usepackage{algorithm}
\usepackage{algorithmic}
\usepackage{enumitem}
\usepackage{url}
\usepackage{xcolor}
\usepackage{acmart-taps}

\newcommand\algorithmicfunction{\textbf{function}}
\renewcommand\algorithmicrequire{\textbf{Input:}}
\renewcommand\algorithmicensure{\textbf{Output:}}
\newcommand\textproc{\textsc}
\newcommand\Require{\item[\algorithmicrequire]}%
\newcommand\Ensure{\item[\algorithmicensure]}%
\newcommand\Call[2]{\textproc{#1}\ifthenelse{\equal{#2}{}}{}{(#2)}}
\newcommand\Function[2]{\State\algorithmicfunction\ \textproc{#1}\ifthenelse{\equal{#2}{}}{}{(#2)}\begin{ALC@g}}
\newcommand\EndFunction{\end{ALC@g}\State\algorithmicend\ \algorithmicfunction}
\newcommand\State\STATE
\newcommand\For\FOR
\newcommand\EndFor\ENDFOR
\newcommand\While\WHILE
\newcommand\EndWhile\ENDWHILE

\usepackage{tikz}
\usetikzlibrary{fit,calc}

\usepackage{booktabs} 
\usepackage{caption}
\usepackage{tabularx}

\newcommand\OurMethodName{SLOPT}
\newcommand\HavocMAB{\textit{Havoc}\textsubscript{\textit{MAB}}}

\newcommand*{\tikzmk}[1]{\tikz[remember picture,overlay,] \node (#1) {};\ignorespaces}

\newcommand{\boxita}[1]{\tikz[remember picture,overlay]{\node[yshift=1pt,fill=#1,opacity=.15,fit={($(A)-(.11\linewidth,0)$)($(B)+(.25\linewidth,.3\baselineskip)$)}] {};}\ignorespaces}
\newcommand{\boxitb}[1]{\tikz[remember picture,overlay]{\node[yshift=1pt,fill=#1,opacity=.20,fit={($(A)-(.05\linewidth,0)$)($(B)+(.30\linewidth,.3\baselineskip)$)}] {};}\ignorespaces}

\colorlet{pink}{red!40}
\colorlet{lettuce}{green!50}

\begin{document}

\setlength\textfloatsep{3mm}
\setlength\floatsep{3mm}
\setlength\intextsep{1pt}

\title{\OurMethodName{}: Bandit Optimization Framework for Mutation-Based Fuzzing}

\author{Yuki Koike}
\email{yukik@ricsec.co.jp}
\affiliation{%
  \institution{Ricerca Security, Inc.}
  \country{}
}

\author{Hiroyuki Katsura}
\email{hiroyukik@ricsec.co.jp}
\affiliation{%
  \institution{Ricerca Security, Inc.}
  \country{}
}

\author{Hiromu Yakura}
\email{hiromu.yakura@aist.go.jp}
\affiliation{%
  \institution{University of Tsukuba / National Institute of Advanced Industrial Science and Technology (AIST)}
  \country{}
}

\author{Yuma Kurogome}
\email{yumak@ricsec.co.jp}
\affiliation{%
  \institution{Ricerca Security, Inc.}
  \country{}
}

\renewcommand{\shortauthors}{Koike et al.}

\begin{abstract}
Mutation-based fuzzing has become one of the most common vulnerability discovery solutions over the last decade.
Fuzzing can be optimized when targeting specific programs, and given that, some studies have employed online optimization methods to do it automatically, i.e., tuning fuzzers for any given program in a program-agnostic manner.
However, previous studies have neither fully explored mutation schemes suitable for online optimization methods, nor online optimization methods suitable for mutation schemes.
In this study, we propose an optimization framework called \OurMethodName{} that encompasses both a \textit{bandit-friendly} mutation scheme and \textit{mutation-scheme-friendly} bandit algorithms.
The advantage of \OurMethodName{} is that it can generally be incorporated into existing fuzzers, such as AFL and Honggfuzz.
As a proof of concept, we implemented \OurMethodName-AFL++ by integrating \OurMethodName{} into AFL++ and showed that the program-agnostic optimization delivered by \OurMethodName{} enabled \OurMethodName-AFL++ to achieve higher code coverage than AFL++ in all of ten real-world FuzzBench programs.
Moreover, we ran \OurMethodName-AFL++ against several real-world programs from OSS-Fuzz and successfully identified three previously unknown vulnerabilities, even though these programs have been fuzzed by AFL++ for a considerable number of CPU days on OSS-Fuzz.
\end{abstract}




\maketitle

\section{Introduction}

Fuzzing is one of the most effective and widely-used methods for identifying vulnerabilities. 
Given a targeted program under test (PUT), the goal of fuzzing is to find inputs that cause abnormal termination of the program.
To address this problem, fuzzers iteratively mutate a given set of test cases or generate random byte sequences with a given grammar to obtain new inputs and execute the program with those inputs.
Despite the simplicity of its mechanism, fuzzing has discovered a large number of bugs and vulnerabilities and has contributed considerably to improving the quality and security of software \cite{SAGE, ClusterFuzz}. 

Fuzzing can be customized for specific types of PUT to maximize its effectiveness.
For example, extensive research has been conducted on the design of fuzzers targeting file formats with a fixed syntax, such as XML, and software with semantics, such as JavaScript engines and database management systems \cite{NAUTILUS, DIE, FuzzIL, SQUIRREL, syzkaller}.
To be more general, fuzzers can be optimized if users have a clear scope of programs to fuzz and can devote engineering efforts to make them reflect the characteristics of the targeted programs, as in structure-aware \cite{AFLSmart, protobufmutator} and human-in-the-loop fuzzing \cite{IJON}.

On the other hand, fuzzers have already been used to ensure software quality and security in an unspecified large number of software development projects, as seen in OSS-Fuzz \cite{OSSFuzz}.
In most cases, fuzzers are considered just one of the automated services used during development, and it is possible that developers cannot afford the cost of customizing and effectively using fuzzers.
This suggests that it is desirable to automatically tune fuzzers by considering the characteristics of the provided programs to maximize their effectiveness without requiring additional engineering efforts.

To meet this demand, online optimization algorithms, particularly bandit algorithms, have gained attention in recent years as a means of accelerating the efficiency of mutation-based fuzzing \cite{ScheduleBlackBox, EcoFuzz, SyzVegas, AdaptiveFuzzTS, CMFuzz, HierarSeed, SIVO, HavocMAB}.
They are expected to work with a minimum amount of information that can be retrieved in any environment as long as feedback-driven fuzzing can be used. 
For example, mutation-based fuzzing generates new input by randomly making several choices, including the choice of seeds to be mutated and the mutation method.
By treating these choices as bandit problems, fuzzers have successfully found more execution paths, and thus, bugs.

In this study, we propose \OurMethodName, a new optimization framework that uses bandit algorithms to optimize two particular choices in mutation-based fuzzing: the method of mutation and the number of mutations to be performed on a seed at once.
This allows fuzzers to distinguish which method and number of mutations are more likely to discover new execution paths at runtime without any prior knowledge.
To take full advantage of bandit algorithms, we made the following three key observations upon incorporating them.

\noindent\textbf{Bandit-friendly mutation scheme}\ \ We introduce a new mutation scheme that is, so to speak, \textit{bandit-friendly} so that the two choices in mutation can be easily and directly treated as bandit problems.
Whereas the typical mutation scheme used in existing fuzzers applies multiple types of mutation operators at a time, our scheme uses exactly one type of mutation operator per mutation, considering that only one arm can be selected per step in a bandit problem.
We experimentally show that a fuzzer achieves similar code coverage regardless of whether it follows the conventional scheme or our bandit-friendly scheme, which allows the modification of existing fuzzers so that they follow the bandit-friendly scheme.
This mutation scheme resolves the dependency of existing methods on heuristics that were introduced to harness online optimization without changing the mutation scheme \cite{AdaptiveFuzzTS, CMFuzz, MOpt}.
Although these heuristics work well in practice, it is also beneficial to adhere to vanilla optimization algorithms by changing the scheme itself; this enables us to take advantage of the theoretical guarantees and advancements in the well-researched field of optimization algorithms. For example, it enables the direct incorporation of unseen algorithms to be proposed in the near future into fuzzing.

\noindent\textbf{Optimization of batch size}\ \ Our optimization target includes not only the mutation operators but also the number of times mutations are applied to one seed at a time, referred to hereafter as \textit{batch size}.
We acknowledge that this target had been already employed in the existing method \HavocMAB{} \cite{HavocMAB}. This was based on the observation that a fuzzer showed different performances with batch size set to different fixed values. However, we focus on this from another observation and perspective, which consequently differentiates \HavocMAB{} from our resultant optimization method.
Specifically, we found that the optimal batch size can vary depending on the length of the seed to be mutated; therefore, \OurMethodName{} considers the lengths of seeds when optimizing the batch size.

\noindent\textbf{Bandit algorithms suitable for fuzzing}\ \ We further conducted a comprehensive comparison of bandit algorithms and determined the most suitable one for our framework.
Although there are various types of bandit algorithms, the extent to which each type of algorithm works effectively with fuzzing has not been investigated in detail, possibly because the aforementioned heuristics have hindered the adoption of various algorithms.
Hence, we ran a thorough experiment to compare classical, non-stationary, and adversarial bandit algorithms by incorporating them into a fuzzer and measuring their performance.
We then found that the fuzzers with adversarial bandit algorithms performed worse than those with other algorithms in our mutation scheme.

With these key observations, as a proof of concept, we implemented \OurMethodName-AFL++ on top of AFL++ \cite{AFLpp} and measured its performance extensively on two benchmarks: FuzzBench~\cite{fuzzbench} and MAGMA~\cite{MAGMA}.
In every PUT, \OurMethodName-AFL++ achieved a higher code coverage than its baseline, AFL++, proving that it realizes PUT-agnostic optimization.
We also compared the performance improvement achieved by \OurMethodName{} with that of the existing online optimization methods.
In the comparison, we observed that \OurMethodName{} outperformed the others on average in the tested PUTs and that each optimization method provided various levels of improvement for different PUTs, which indicates the future potential of our approach by combining with them. 

Additionally, we ran \OurMethodName-AFL++ against several PUTs in OSS-Fuzz for seven days to determine whether it could find crashes and vulnerabilities in real-world programs.
As a result, \OurMethodName-AFL++ found 17 unique bugs, three of which were confirmed as previously undiscovered vulnerabilities and assigned CVE IDs.
We released \OurMethodName-AFL++ on \url{https://github.com/RICSecLab/SLOPTAFLpp}.

\section{Background}

\subsection{Mutation-Based Fuzzing}

Given a target program (program under test; PUT) and a set of initial test cases for the program, mutation-based fuzzers run in a semi-permanent loop, generating inputs and feeding them to the program.
In particular, in each iteration of the loop, the fuzzers perform the following steps.

\noindent\textbf{Seed scheduling}\ \ First, the fuzzers choose a source test case to generate a new input and the number of inputs to be generated.
In fuzzing, the test cases are called seeds because the chosen test case works as a sample of the new inputs, and the policy of selecting seeds is called \textit{seed scheduling}.
Common seed-scheduling methods include selecting seeds from a seed queue in order or at random.

\noindent\textbf{Mutation}\ \ The fuzzers then modify the chosen seed the determined number of times to create a new input.
In many cases, modifications made to the seed are selected from a predetermined set of operations, such as bit flipping or insertion of a constant byte sequence.
These operations are known as \textit{mutation operators}. 
Usually, fuzzers also decide which position in the input to apply the mutation operator and the other arguments required by the mutation operator.

\noindent\textbf{Execution and Update}\ \ After generating an input, the fuzzers execute the PUT and obtain feedback as a result.
In greybox fuzzing, for example, they can acquire code coverage, by performing lightweight instrumentation on the PUT.
If they find the new input valuable through feedback, they save it as a seed.
When the fuzzers employ dynamic optimizations in their scheduling or mutation, they then update the components based on feedback.

Thus, mutation-based fuzzers have many choices to make in their fuzzing process, which affect the efficiency of the fuzzers.

\subsection{Multi-Armed Bandit Problem}

The multi-armed bandit problem is analogous to playing slot machines, whose goal is to find the most profitable arm (choice) from a fixed set of multiple arms and keep pulling (selecting) it as many times as possible.
Formally, a decision maker is given $K$ probability distributions over real numbers, and in each round, the decision maker chooses one of the distributions and is rewarded with a real number sampled from that distribution.
The goal is to maximize the sum of rewards obtained as much as possible over $T$ rounds, without knowing the parameters of the $K$ probability distributions.

Despite the simplicity of its problem formulation, there are many variations; in the most basic variation (i.e., a regular stochastic bandit problem), the $K$ probability distributions are defined as independent Bernoulli distributions.
However, they can be substituted by other distributions such as normal distributions in other settings.
The non-stationary stochastic bandit problem assumes the probability distributions of the rewards to change with time.
The adversarial bandit problem assumes the most extreme situation, where the rewards given by the arms in $T$ rounds are arbitrary and not assumed to follow any probability distribution.

In the study of the application of bandit algorithms to fuzzing, rewards are often assumed to take only the discrete value of $0$ or $1$.
This is because the plausible rewards settable in fuzzing are whether a new crash, code block, or execution path is found by executing a PUT.
Our study is no exception: we use the discovery of execution paths as a reward.
Note that our proposed method is valid even if the reward is replaced by the discovery of crashes or code blocks.

\section{Related Work}

\subsection{Improvements on Mutation-Based Fuzzer}

Various components of mutation-based fuzzers have been refined from different perspectives to improve their efficiency, that is, increase code coverage, particularly after the advent of the representative feedback-driven fuzzer AFL \cite{AFL}.
Some studies achieved efficiency purely by speeding up the execution of PUTs and the acquisition of feedback from that execution \cite{forkserver, PTfuzz, PTrix, UnTracer}.
Other studies aimed to generate inputs that are more likely to find new code blocks by supplying extra information into feedback with binary instrumentation, taint inference, or emulators, and/or by integrating other techniques to solve constraints such as the SMT solver \cite{VUzzer, REDQUEEN, Eclipser, Angora, KLEE, Driller, QSym}.
However, these technologies often have non-negligible additional costs to deploy; they usually reduce the execution speed of PUTs and thus lower the throughput of a fuzzer.
They may also support a limited range of program execution environments, narrowing the range of PUTs that can be fuzzed.

Because of this concern, a series of studies have explored optimizations to make existing feedback-based fuzzers work efficiently against any PUTs without additional information.
Most studies focused on either seed selection or mutation because they are essential parts of mutation-based fuzzers, which simply repeat the loop of selecting a seed, modifying it to create new inputs, and executing a PUT with them.
For example, AFLFast optimized a parameter called \textit{energy}, which denotes the number of times a seed is used to generate new inputs by modeling fuzzing as a Markov chain \cite{AFLFast}.
Similarly, B\"{o}hme et al.\ employed information theory \cite{Entropic} and Patil and Kanade adopted reinforcement learning \cite{GreyFuzzAsContextual} to regulate the energy parameter.
In another direction, Rebert et al.\ optimized seeds to be kept in a seed set as a minimal set cover problem \cite{OptimizSeedSelection}. 
FairFuzz introduced a mutation mask to limit the positions in the input to be modified, thereby increasing the likelihood that new inputs will trigger rare execution paths \cite{FairFuzz}.
Our study has the same motivation as these studies to achieve PUT-agnostic optimization; however, we focus on applying bandit algorithms because they can be further leveraged by introducing our bandit-friendly scheme.

\subsection{Online Optimization of Mutation Operator}

When we consider applying online optimization to fuzzing, there are several options for selecting optimization targets.
Although some studies have employed bandit problems to optimize seed scheduling and other components in fuzzing \cite{ScheduleBlackBox, EcoFuzz, HierarSeed}, previous studies applying online optimization to the selection of mutation operators are the most relevant to our work.
For example, MOpt optimized the probability distribution of choosing mutation operators during runtime with its original heuristics that resemble Particle Swarm Optimization, based on the observation that different PUTs have different mutation operators that are likely to produce interesting seeds \cite{MOpt}.
However, as shown in Section~\ref{sec:evaluation}, MOpt often results in worse performance than before the optimization and therefore can be unsatisfactory as an online method for optimizing the distribution in a PUT-agnostic manner.

Karamcheti et al.\ integrated Thompson sampling into a random mutation in AFL, called havoc, and confirmed its effectiveness \cite{AdaptiveFuzzTS}.
Although they modified only AFL, their mutation scheme is generally applicable to other fuzzers.
However, unlike our scheme, their scheme is troubled by the credit assignment problem.
They attached Thompson sampling to their scheme without modifying the havoc, resulting in the need to select multiple mutation operators in one step.
In principle, this is clearly intractable with classical stochastic bandit problems because the algorithms can select only one arm in one step in those problems.
They bypassed this issue by heuristically fixing the number of mutation operators to be selected at a time (referred to as \texttt{sample\_num\_mutations}, hereafter called \textit{batch size}).
This leaves an open problem regarding the application of bandit algorithms with arbitrary batch size.
In Section~\ref{sec:mutationscheme}, we show that by slightly changing the traditional scheme that havoc follows, we can avoid the credit assignment problem and solve this open problem.

CMFuzz extended the mutation scheme of Karamcheti et al.\ by treating the seeds to be mutated as a context \cite{CMFuzz}. 
It then modeled the choice of the mutation operator as a contextual bandit problem and used LinUCB to optimize it.
This is promising because the contextual bandit problem can consider the differences between seeds.
However, like the scheme of Karamcheti et al., the mutation scheme of CMFuzz is also affected by the credit assignment problem; thus, the batch size must be set to a small constant.

Wu et al. empirically examined the characteristics of havoc and sublimed their discoveries into a new mutation scheme that allows bandit algorithms to adaptively optimize the selection of mutation operators and batch sizes \cite{HavocMAB}. While their proposed algorithm, \HavocMAB{}, is the most similar to ours, there are two major differences. First, \HavocMAB{} uses a single instance of the bandit problem to optimize a parameter, whereas our algorithm chooses one from multiple instances depending on the length of the seed to be mutated. Our design is based on the observation that the optimal choice of parameters is conditioned by length, as described in Section~\ref{sec:newtarget}. Second, unlike ours, \HavocMAB{} simultaneously applies different types of mutation operators to a single input, similar to the algorithms of Karamcheti et al. and CMFuzz. With such algorithms, it is difficult to analyze which mutation operator is effective when a generated input increases the code coverage, which can hinder fuzzers from making precise choices.

SIVO introduced various contrivances, such as taint inference engine, symbolic constraint solver, and customized instrumentation, and optimized all optimizable parameters using a bandit algorithm \cite{SIVO}.
Notably, it employed non-stationary bandit algorithms to mitigate the fact that, in most cases, rewards monotonically decrease when fuzzing is formulated as a bandit problem.
Still, the bandit algorithm employed in SIVO is discounted Boltzmann exploration, which was originally invented by SIVO, to the best of our knowledge, by applying the technique of discounting \cite{dUCB, dTS} to Boltzmann exploration.
As discussed in Section~\ref{sec:banditcomparison}, there can be room for further performance improvement by adopting other non-stationary bandit algorithms.

\section{Overview of \OurMethodName{}}
\label{sec:overview}
\subsection{Bandit-Friendly Mutation Scheme}
\label{sec:mutationscheme}

\aptLtoX[graphic=no,type=env]{\begin{center}
\includegraphics{figure/Algorithm1.pdf}
\end{center}
\begin{center}
\includegraphics{figure/Algorithm2.pdf}
\end{center}}{\begin{figure*}
\begin{minipage}{\columnwidth}
\begin{algorithm}[H]
\centering
\caption{Conventional Random Mutation Scheme}
\label{alg:conv}
\begin{algorithmic}[0]
\vspace{-0.15\baselineskip}
\Require{$seed$ -- a test case to be mutated}
\vspace{-0.15\baselineskip}
\Ensure{$input$ -- a new input to be tested}
\vspace{5pt}
\Function{RandomMutation}{$seed$}
\vspace{-0.15\baselineskip}
\State $input$ $\gets$ \Call{CopyBytesFromSeed}{$seed$}
\vspace{-0.15\baselineskip}
\State $batch\_size$ $\gets$ \Call{DecideBatchSize}{\null}
\vspace{-0.15\baselineskip}
\For{$i$ $\gets$ $1$ \textbf{to} $batch\_size$}
\vspace{-0.15\baselineskip}
\State 
    \tikzmk{A}
    $mutation$ $\gets$ \Call{SelectOperator}{\null}
    \tikzmk{B}
    \boxita{red}
    \vspace{-0.15\baselineskip}
    \State $pos$ $\gets$ \Call{SelectPosition}{$input$}
    \vspace{-0.15\baselineskip}
    \State $input$ $\gets$ \Call{ApplyOperator}{$mutation, input, pos$}
\vspace{-0.15\baselineskip}
\EndFor
\vspace{-0.15\baselineskip}
\State \textbf{return} $input$
\vspace{-0.15\baselineskip}
\EndFunction
\end{algorithmic}
\end{algorithm}
\end{minipage}
\hfill
\begin{minipage}{\columnwidth}
\begin{algorithm}[H]
\centering
\caption{Our Random Mutation Scheme}
\label{alg:ours}
\begin{algorithmic}[0]
\vspace{-0.15\baselineskip}
\Require{$seed$ -- a test case to be mutated}
\vspace{-0.15\baselineskip}
\Ensure{$input$ -- a new input to be tested}
\vspace{5pt}
\Function{RandomMutation}{$seed$}
\vspace{-0.15\baselineskip}
\State $input$ $\gets$ \Call{CopyBytesFromSeed}{$seed$}
\vspace{-0.15\baselineskip}
\State $batch\_size$ $\gets$ \Call{DecideBatchSize}{\null}
\vspace{-0.15\baselineskip}
\State 
       \tikzmk{A}
       $mutation$ $\gets$ \Call{SelectOperator}{\null}
       \tikzmk{B}
       \boxitb{lettuce}
\vspace{-0.15\baselineskip}
\For{$i$ $\gets$ $1$ \textbf{to} $batch\_size$}
\vspace{-0.15\baselineskip}
    \State $pos$ $\gets$ \Call{SelectPosition}{$input$}
\vspace{-0.15\baselineskip}
    \State $input$ $\gets$ \Call{ApplyOperator}{$mutation, input, pos$}
\vspace{-0.15\baselineskip}
\EndFor
\vspace{-0.15\baselineskip}
\State \textbf{return} $input$
\vspace{-0.15\baselineskip}
\EndFunction
\end{algorithmic}
\end{algorithm}
\end{minipage}
\end{figure*}}

Algorithm~\ref{alg:conv} shows a scheme of random mutation that modifies a given seed, commonly observed in existing fuzzers including the AFL family (e.g., \cite{AFL, AFLFast, lafintel, AFLpp}), Honggfuzz \cite{honggfuzz}, and Angora \cite{Angora}. 
This mutation scheme selects a mutation operator and applies it to a source seed as many times as the value of $batch\_size$.
As a result, even if the created input finds a new execution path in a PUT (that is, obtains a reward), it becomes unclear which one has contributed to the discovery among the different selected operators.
Thus, the scheme does not have a favorable form for optimizing the probability distribution of selecting mutation operators.

We have reconstructed this scheme based on the following assumption: the multiple mutation operators used to create a single input are unlikely to work together. In other words, the scheme only applies several mostly independent mutation operators simultaneously and tests them on a PUT at once.
Thus, the scheme can be considered as a type of batch processing in which mutation operators are applied in batches.

\begin{figure}[tb]
\centering
\includegraphics[width=.33\textwidth]{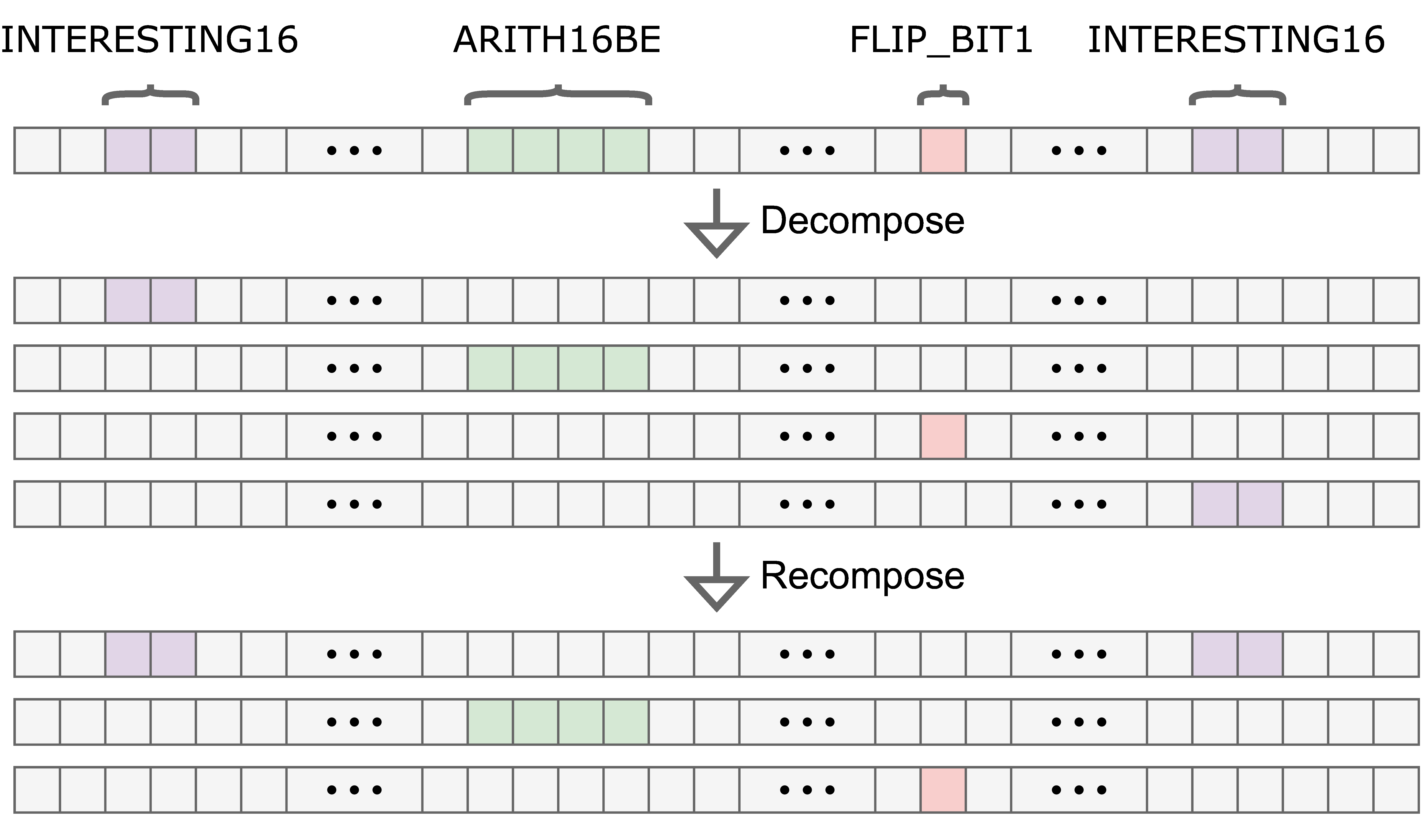}
\caption{Illustration of the essential concept of Algorithm~\ref{alg:ours}.
Each square represents a byte in the input, and colored squares indicate bytes that have been modified by mutation.}
\label{fig:renovate_havoc}
\end{figure}

Figure~\ref{fig:renovate_havoc} describes the essential concept of the proposed scheme shown in Algorithm~\ref{alg:ours}, which originates from our assumption.
Under this assumption, we do not necessarily need to apply multiple mutation operators to a single input.
In particular, we would be able to obtain the same code coverage by applying each mutation operator to different copies of the input and testing them separately.
Furthermore, the assumption allows us to recombine these separated mutation operators as we like because the operators do not affect each other.
This justifies the inclusion of only one type of mutation operator per batch in Algorithm~\ref{alg:ours}.

However, we would not be able to say that the assumption is universally applicable.
For instance, if a large number of mutation operators are included in a single batch, it is extremely likely that a PUT will find errors in the created input at a very early stage of its execution, resulting in no new execution path.
If we did not pack so many mutation operators into a batch, and instead applied the operators to different copies of the input, we would have been more likely to find a new execution path.
Hence, in this case, we can say the mutation operators are mutually dependent.

Therefore, to ensure that our assumption can be applied \textbf{in most cases} and can serve as a basis for our approach, we incorporated these two schemes into AFL++ and compared the quantity of code coverage obtained.
If these two versions of AFL++ have no significant difference in code coverage, we can say that our assumption is valid in practice, which eventually means Algorithm~\ref{alg:ours} is practicable.
Note that our assumption does not need to be (and would not be) valid \textbf{always}. The important thing here is whether fuzzers should persist in the conventional scheme Algorithm~\ref{alg:conv} at all costs. 

We ran the two versions for 24 h against 10 PUTs selected from FuzzBench ten times each.
The method for selecting PUTs and other detailed experimental setups are described in Section~\ref{subsec:eval-setup}.
Note that AFL++ follows Algorithm~\ref{alg:conv} from the beginning. 

\begin{figure*}
\centering
\includegraphics[width=0.93\linewidth]{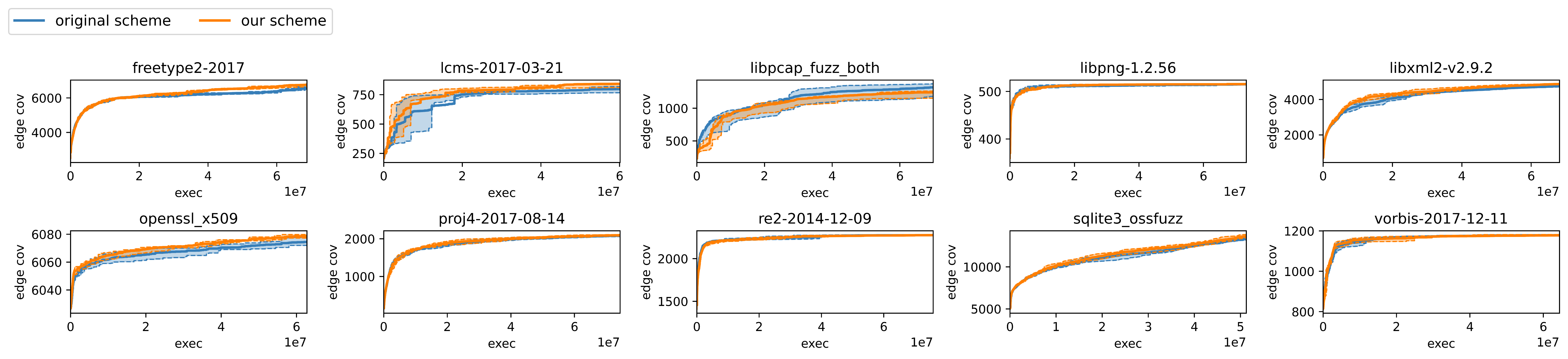}
\caption{Comparison between Algorithm~\ref{alg:conv} and Algorithm~\ref{alg:ours}. The x-axis and y-axis show the number of generated inputs and edges covered, respectively. The solid line, and upper and lower dashed lines show 50\%, 75\% and 25\% quantile, respectively.}
\label{fig:comp_scheme_with_coarse}
\end{figure*}

The results are shown in Figure~\ref{fig:comp_scheme_with_coarse}.
In all the tested PUTs, the AFL++ incorporating our scheme did not show any serious deterioration or stagnation compared to the original AFL++.
In the figure, the two versions exhibit similar increasing trends, and the AFL++ incorporating our scheme achieves even better code coverage in some PUTs.
In other words, we observe no fatal drawbacks when employing Algorithm~\ref{alg:ours} instead of Algorithm~\ref{alg:conv}.
This allows us to reasonably convert fuzzers following Algorithm~\ref{alg:conv} into those following Algorithm~\ref{alg:ours} to directly apply the bandit algorithms.

\subsection{Optimizing Batch Size}
\label{sec:newtarget}

Given a seed, mutation-based fuzzers modify it by applying mutations several times.
In the case of AFL, the number of mutations applied to a seed is determined by \( 2^t \) where \( t \) is a random number. 
We denote \( t \) as the \emph{batch exponent} and \( 2^t \) as the \emph{batch size}.  

Wu et al.\ suggested that optimizing the batch exponent for each PUT could accelerate fuzzers \cite{HavocMAB}.
We also focus on the batch size but our idea is different in that when we optimize the batch size, we consider not only the difference of PUTs but also 
the size of the seed to be mutated and the mutation method to be applied.

\begin{figure*}
  \centering
  \includegraphics[width=0.9\linewidth]{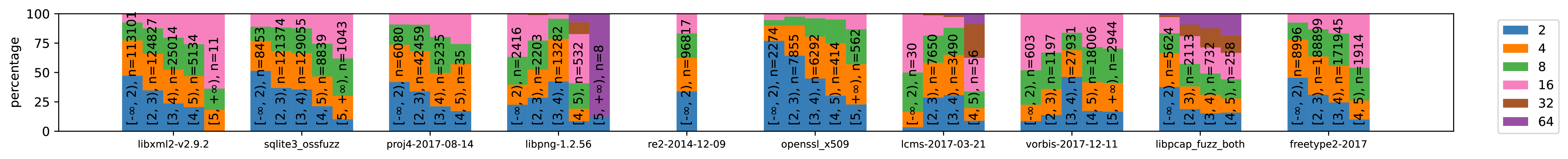}
  \caption{Ratio between batch sizes of the number of seeds saved during a fuzzing campaign by AFL++ for different PUTs and different groups of seed sizes. The label `[x, y)' in a bar means that this bar is the summary of the ratio for the seeds whose size is between \( 2^x \) and \( 2^y \). The number specified by \( n \) is how many seeds are in this bucket as a whole.}
  \label{fig:reptition-observations}
\end{figure*}

First, we hypothesize that the best batch size is dependent on the seed size.
To validate this hypothesis, we collected seeds saved during fuzzing campaigns with AFL++\footnote{AFL++ saves a seed when an input using the seed discovers a new execution path.}, grouped them by the sizes of their parent seeds, and analyzed how many times mutations were applied to produce them.
We executed AFL++ 30 times against PUTs taken from FuzzBench for 24h, and the results are summarized in Figure~\ref{fig:reptition-observations}.
Even though AFL++ chooses batch sizes randomly
\footnote{To be precise, batch exponents greater than 4 are not candidates until AFL++ find no new execution path for a long time.
However, this heuristic would not spoil the discussion because it is still random enough and we actually see different distributions.
},
the distributions of effective batch sizes are different, depending on seed sizes even in the same PUT.
This implies that, rather than assigning the same probability to all batch sizes without considering seed sizes, we should weight them appropriately based on the sizes to find a new execution path efficiently.
Hence, we prepare different bandit problem instances for different seed sizes.

As a side note, it is interesting that, in half of the PUTs, as the seed size increases, the ratio of batch size \( 2^2 \) decreases, which indicates that the possibility of triggering a new execution path with \( 2^2 \) times of mutations decreases as the seed to be mutated becomes larger. 
This phenomenon may be interpreted from the perspective of \textit{batch processing} as explained in Section~\ref{sec:mutationscheme}; the bigger the number of mutation operators applied in parallel gets, the more efficient the random mutation becomes.
On the other hand, mutating too many positions of a seed can easily break the whole structure of the seed by destroying its important parts (e.g., magic number).
Larger seed size progressively lowers such possibility by relatively decreasing the probability of the important parts being selected as mutated positions.
However, this is just a possible explanation and this phenomenon requires further investigation, considering that it is not applicable to the other half of the PUTs.

In addition, when optimizing the batch size, we should be aware of the differences in mutation operators selected.
Various mutation operators have been developed to find new code blocks efficiently. 
For example, AFL-based fuzzers contain the mutation operators named `flip\_bit' and `clone\_bytes';
the former only modifies one byte of the original seed while the latter appends a possibly great number of bytes to it.
To precisely control the number of positions to be modified, we should prepare different bandit problem instances for each mutation operator.

Based on these viewpoints, we designed \OurMethodName{} to determine the batch exponent using the following mechanism: instead of using one fixed batch exponent or a random number, we introduce our bandit algorithm suitable for fuzzing, as explained in Section~\ref{sec:banditcomparison}, to optimize the batch exponent for each PUT.
We prepared different bandit problem instances for different seed sizes and mutation operators. 
Because the seed sizes can be arbitrary non-negative numbers and are too many to prepare different bandit instances for each of them,  
we prepare five groups for seeds with sizes of \( [0, 10^2) \), \( [10^2, 10^3) \), \( [10^3, 10^4) \), \( [10^4, 10^5) \), and \( [10^5, \infty) \). 
Even though there is room for tuning, our observations indicated that these groups work well for capturing the overall tendency of seed sizes.

Because a batch exponent in the original AFL++ can only be at most seven, we prepare seven bandit arms for each instance; the \( t \)-th arm corresponds to the batch exponent \( t \).
Given a source seed to be mutated and a mutation operator to be applied to the seed, \OurMethodName{} fetches the bandit instance corresponding to the seed size and the mutation operator.
Then, it pulls an arm of the bandit instance to obtain the batch exponent \( t \).
After that, it mutates the seed \( 2^t \) times and tests if the execution with the mutated seed discovers a new execution path.
If so, it gives a reward to the arm.

\subsection{Bandit Algorithms for Fuzzing}
\label{sec:banditcomparison}	

Although our scheme is designed to accommodate bandit algorithms naturally, not all bandit algorithms may yield the same level of performance improvement.
As previous studies have highlighted, if fuzzing is viewed as a bandit problem and reward is defined as the discovery of a new execution path or crash, the rewards obtained will decrease with time because the discovery can only be made a finite number of times \cite{ScheduleBlackBox, EcoFuzz, SyzVegas, SIVO}.
As a result, the estimates of the expected rewards become poor. This can lead to worse performance in some bandit algorithms, such as UCB1 \cite{UCB1}, because they rely on accurate estimates of the expected rewards to select a promising arm an appropriate number of times; the poor estimates themselves may deteriorate the algorithm performance even if the best arm is still correctly inferred.
Thus, these previous studies employed bandit algorithms that tolerate changes in expected rewards, such as adversarial or non-stationary bandit algorithms.

\begin{table}[tb]
\centering
\caption{List of bandit algorithms implemented.}
\label{tab:bandit_algorithms}
\begin{tabular}{l}
\toprule
Classical (Stationary) Bandit Algorithm:                             \\
\midrule
\, UCB1 \cite{UCB1} \, / \, KL-UCB \cite{KLUCB}                      \\
\, Thompson sampling (TS) \cite{TS}                                  \\
\cmidrule[1pt]{1-1}
Non-Stationary Bandit Algorithm:                                     \\
\midrule
\, discounted TS (dTS) \cite{dTS}                                    \\
\, discounted Boltzmann Exploration (dBE) \cite{BE, SIVO, SIVOimpl}  \\
\, adaptive shrinking TS (ADS-TS) \!\cite{ScaleMAB, ADS}             \\
\cmidrule[1pt]{1-1}
Adversarial Bandit Algorithm:                                        \\
\midrule
\, EXP3-IX \cite{Exp3ix} \, / \, EXP3++ \cite{Exp3pp, ImpParaExp3pp} \\
\bottomrule
\end{tabular}
\end{table}

\begin{table}[tb]
\centering
\caption{Ranks and scores averaged over 10 PUTs of AFL++ and 8 versions of \OurMethodName-AFL++ after 24 h. }

\begin{tabular}{lccccc}
\toprule

& AFL++ & UCB1 & KL-UCB & TS & dTS \\
\midrule

Rank Avg & \textit{ 7.9 } & 5.5 & 3.5 & 3.9 & 3.0 \\
Score Avg & \textit{ 89.44 } & 93.83 & 96.58 & \textbf{ 98.81 } & 97.85 \\

\midrule

& dBE & ADS-TS & EXP3-IX & EXP3++ \\
\midrule

Rank Avg & 4.9 & \textbf{ 2.6 } & 6.5 & 6.9 \\
Score Avg & 95.57 & 97.39 & 94.11 & 93.56 \\

\bottomrule

\end{tabular}

\label{tab:alg_cmp_summary}
\end{table}

To check whether this concern seriously affects the performance and to determine which bandit algorithm should be used in \OurMethodName{}, we implemented several versions of \OurMethodName-AFL++ with various bandit algorithms and ran them against 10 PUTs from OSS-Fuzz~\cite{OSSFuzz} for 24 h ten times each. 
We prepared PUTs from OSS-Fuzz because, as advised by B\"{o}hme et al. \cite{FuzzReliability}, we needed to avoid overfitting to the benchmarks in Section~\ref{sec:evaluation} by knowing which algorithm works best in PUTs of the benchmarks. However, simultaneously, we also wanted to make this preliminary comparison further applicable by carrying it out with a diverse variety of real-world programs.
For the same purpose, we made the configuration of each PUT, including initial seeds, dictionaries, and command line arguments fed to the fuzzers, consistent with that used in the OSS-Fuzz infrastructure. 
Moreover, to minimize selection bias, we first enumerated 64 projects in OSS-Fuzz that met certain requirements and then selected 10 projects randomly among them (listed in Table~\ref{tab:alg_cmp_all}).
The requirements we posed were that the project must be compatible with AFL++, and that the project must be built with no error for the sake of avoiding experimenter bias introduced by ad-hoc fixes at our own discretion.

Table~\ref{tab:bandit_algorithms} lists bandit algorithms we implemented.
Note that regarding the implementation of dBE, we completely followed SIVO \cite{SIVO} by using its configuration of hyperparameters and introducing its heuristics, although its implementation of Boltzmann exploration diverged from those in literature owning to the heuristics \cite{SIVOimpl}.
Other details of the experimental settings are provided in Section~\ref{subsec:eval-setup}. 

Table~\ref{tab:alg_cmp_summary} represents the ranks and scores of eight versions of \OurMethodName-AFL++ averaged over the 10 PUTs, which can also be seen in the reports of FuzzBench \cite{FuzzbenchReport}.
Both metrics were calculated from the median obtained code coverage\footnote{We define the gain of code coverage as the difference between the initial and final edge coverage. We summarize the initial code coverage of each PUT in Table~\ref{tab:put_details}.} over 10 instances for each PUT. The score of each fuzzer was the proportion of its median coverage to the highest median. 
Among the stationary bandit algorithms, TS provided the best performance improvement on average, whereas UCB1 performed relatively poorly.
Among the non-stationary bandit algorithms, dTS and ADS-TS, which are based on TS, showed a higher performance than dBE and achieved the highest average rank; on the other hand, both adversarial bandit algorithms exhibited noticeably poor performance.
These results are consistent with previous numerical experiments on stochastic bandit problems \cite{UCBEval, TSEval, TSEval2}, which indicates that our mutation scheme can be characterized more as a stochastic bandit problem than an adversarial bandit problem.
Moreover, the non-stationarity of fuzzing as a bandit problem cannot be confirmed in this experiment, as the TS, dTS, and ADS-TS algorithms produced comparable results.
However, this cannot completely disprove the non-stationarity because TS sometimes gives a strong performance even in non-stationary bandit problems, for example, when the best arm is unchanged \cite{TSWinInNonst}.

We must note that, as shown in Table~\ref{tab:alg_cmp_all},  this stochastic tendency can be observed \textbf{on average}, not necessarily on \textbf{each} PUT, and also that there may be different tendencies for different mutation schemes.
Nevertheless, it is likely that fuzzers with bandit optimization can improve their performance by regarding their algorithm as a type of bandit problem and appropriately employing bandit algorithms in that class.
Based on this result, we have adopted TS as the standard algorithm for \OurMethodName-AFL++ for further experiments because it achieved the highest median coverage.

\section{Evaluation}
\label{sec:evaluation}

To demonstrate that our algorithm based on the above observations contributes to the realization of PUT-agnostic online optimization and compare it with existing methods, we evaluated TS-\OurMethodName-AFL++ on major coverage-based and bug-based benchmarks: FuzzBench \cite{fuzzbench} and MAGMA \cite{MAGMA}. 
MAGMA is a ground-truth fuzzing benchmark in real-world software; bugs are inserted artificially and each crash triggered by a fuzzer can be identified.

\subsection{Setup}
\label{subsec:eval-setup}

\noindent\textbf{Experiment Environment}\ \ All experiments, including our initial exploration in Section~\ref{sec:banditcomparison}, were conducted on an NVIDIA DGX A100 640GB workstation with two AMD EPYC 7742 2.25GHz CPUs with 256 logical cores in total.
To avoid fuzzer instances that interfere with each other and produce inaccurate results, we ran only the instances of a single fuzzer in parallel.
Furthermore, to maintain a constant workload, we ensured that at most 110 fuzzer instances were running simultaneously and that the sets of targeted PUTs running together, as well as the number of instances for each PUT among a set, were the same.
Preliminary experiments confirmed that the number of PUT executions per second was not significantly affected even when 110 instances were running simultaneously.

\noindent\textbf{Compared Fuzzers}\ \ We compared (TS-)\OurMethodName-AFL++ with five fuzzers: AFL++, MOpt-AFL++ \cite{MOpt}, AFL++ combined with the algorithm proposed by Karamcheti et al. (hereafter referred to as Karamcheti-AFL++) \cite{AdaptiveFuzzTS}, \HavocMAB{}-AFL++ \cite{HavocMAB}, and CMFuzz-AFL++ \cite{CMFuzz}. The compared fuzzers, including ours, are all based on AFL++ and optimize the selection of mutation operators, except for unaltered AFL++. Notice that only MOpt has been previously integrated in AFL++; therefore, we incorporated other optimization methods into AFL++. Especially, Karamcheti- and CMFuzz-AFL++ were implemented from scratch because their original implementations have not been published.
We did not modify any hyperparameters in each fuzzer if their default values were specified in its implementation. Otherwise, they were set to popular or reasonable values. In particular, for MOpt-AFL++, we set the pacemaker parameter (\texttt{-L}) to 1, noted as the basic choice in \cite{MOptimpl}. For the other details, please refer to our source code. 

\noindent\textbf{PUT Selection}\ \ In FuzzBench, we selected 10 PUTs and ran fuzzers with them to keep the experiment cost reasonable. We first excluded \texttt{jsoncpp\_jsoncpp\_fuzzer}, \texttt{systemd\_fuzz-link-parser}, and \texttt{zlib\_zlib\_uncompress\_fuzzer}, and then randomly selected 10 of the remaining 19 PUTs. We excluded these three PUTs because, as shown in Table~\ref{tab:fuzzbench_max_cov}, even the theoretical maximum values of their code coverage (as well as actual saturation values) are so small that the increase in code coverage would become saturated at a very early stage in every fuzzer; therefore, a comparison with these PUTs would not be meaningful. In fact, this phenomenon was observed in past experiments of FuzzBench \cite{FuzzbenchReport}. While early saturation may also occur in other PUTs even though their code is not small, we did not exclude such PUTs to avoid selection bias caused by excessive filtering. Random selection is also a measure to prevent selection bias. In MAGMA, all PUTs were used because the CPU days required for the evaluation with MAGMA were smaller than that of FuzzBench.
PUT-specific configurations, such as initial seeds and dictionaries, were all unchanged from default.

\noindent\textbf{\#Instances per PUT and Duration}\ \ Although the five fuzzers incorporating optimization methods were expected to improve code coverage and therefore we wanted to contrast them as much as possible especially in FuzzBench, it was concerned that these fuzzers might bring relatively similar results and it could become difficult to compare them by running only a small number of instances of them. Hence, in FuzzBench, we ran 30 instances of each fuzzer for 24 h. In MAGMA, we ran 10 instances of each fuzzer for 24 h. 

\subsection{Effect on Baseline}

\label{subsec:eval-vs-aflpp}

\begin{figure*}[tb]
\centering
\includegraphics[width=0.96\linewidth]{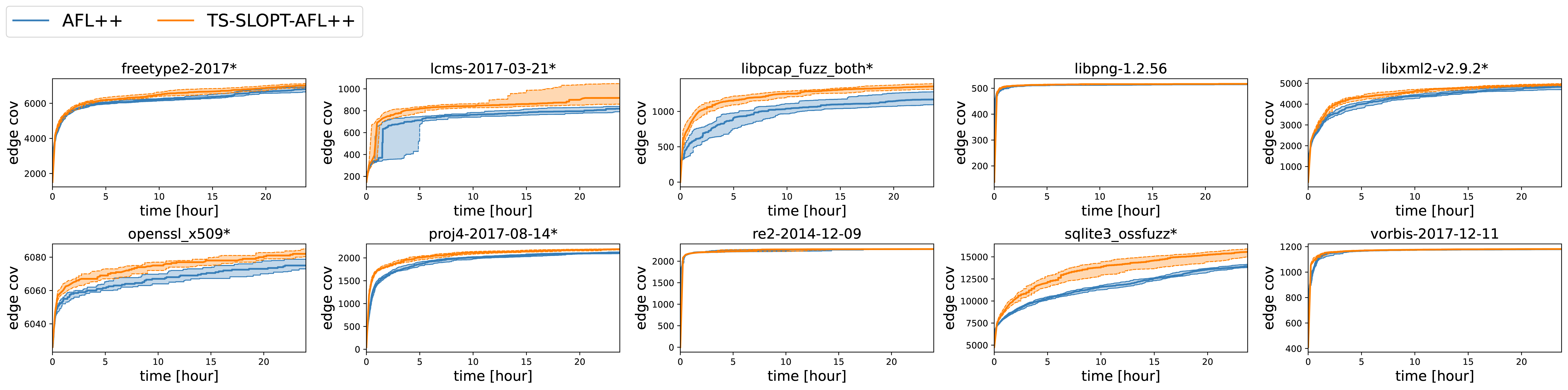}
\caption{Transition of edge coverage of two fuzzers running for 24 h against 10 PUTs taken from FuzzBench. The x-axis represents elapsed hours, and the y-axis shows the edge coverage. The solid line represents the average of the 30 runs. The upper and lower dashed lines show 75\% and 25\% quantile, respectively. The `*' sign to the right of a title indicates that the sets of the edge coverage obtained by the two fuzzers are significantly different in the Mann-Whitney U test ($p < 0.01$). }
\label{fig:edge-cov}
\end{figure*}

We first evaluate the effect of introducing our proposed method.
To this end, we compared TS-\OurMethodName-AFL++ with AFL++ by plotting the code coverage obtained by these fuzzers in FuzzBench.
Because \OurMethodName{} regards the discovery of a new execution path as a reward, the coverage-based benchmark FuzzBench is more suitable than the bug-based benchmark MAGMA to see if \OurMethodName{} works successfully. 
The results are shown in Figure~\ref{fig:edge-cov}. TS-\OurMethodName-AFL++ exhibited faster growth of code coverage within the first few hours for many PUTs, remaining evidently higher than the code coverage of AFL++ until the end. 
Furthermore, for every PUT, the median code coverage of TS-\OurMethodName-AFL++ never fell below that of AFL++ at the end, as presented in Table~\ref{tab:vs_existing}.
This fact is supported statistically by the Mann-Whitney U test.
As shown in Figure~\ref{fig:edge-cov}, the two fuzzers produced significantly different edge coverages in every PUT except the three PUTs in which the fuzzers indistinguishably saturated the code coverage very early.
Thus, we conclude that \OurMethodName{} generally improves performance without any prior knowledge of a PUT.

\subsection{Comparison with Effects of Other Methods}
\label{subsec:eval-vs-existing}

Next, we contrast the performance improvement delivered by \OurMethodName{} with that of existing online optimization methods.
Similar to Section~\ref{subsec:eval-vs-aflpp}, we examined the improvement in code coverage brought about by each method using FuzzBench.
Table~\ref{tab:vs_existing} presents the median edge coverage obtained using the six fuzzers over 24 h.
While all the optimization methods improved the code coverage compared to the baseline AFL++ in some PUTs, MOpt often degraded the code coverage in individual PUTs; hence, its average performance was slightly worse than that of AFL++. In particular, the median obtained coverage of MOpt-AFL++ was 5.78\% lower than that of AFL++ in \texttt{proj4-2017-08-14}.
In contrast, the other optimization methods, all of which incorporate bandit algorithms, presented an edge coverage worse than that of AFL++ by at most 0.36\%.
This suggests that bandit optimization would rarely yield results inferior to uniformly random choices, which is an important attribute for achieving PUT-agnostic optimization.
In particular, in every PUT, \OurMethodName{}-AFL++ was superior to AFL++ when compared by not only the median code coverage but also Vargha-Delaney's $\hat{A}_{12}$ \cite{A12} (or rank sum) as showed in Table~\ref{tab:statistics}. In addition, its performance improvement was the highest in both score and rank on average. 
Conversely, it is notable that some of the existing methods gave results worse than the baseline, which were confirmed as statistically significant in part. MOpt-AFL++ produced such results in four PUTs, as did Karamcheti-AFL++ in \texttt{vorbis-2017-12-11} although there was not a wide margin between its edge coverage and AFL++.

Another interesting point is that the degree of performance improvement brought about by each method is diverse, depending on the PUTs.
For example, Karamcheti-AFL++ achieved the best performance improvement in \texttt{libxml2-v2.9.2} when compared by medians, while \HavocMAB{} in \texttt{libpcap\_fuzz\_both} and ours in \texttt{sqlite3\_ossfuzz}. This indicates that different approaches to online optimization bring about different effects that can be either strong or weak, depending on each PUT, even though they all integrate bandit optimization.
Specifically, the major difference between \OurMethodName{} and the others lies in their mutation schemes, as described in Section~\ref{sec:mutationscheme}, and we believe this difference explains the different degrees of performance improvement between \OurMethodName{} and the others, as described in Section~\ref{subsubsec:case-study}; however, this could not be proven because the experiment with 10 PUTs is still not sufficient as a case study to rule out coincidence.
Nevertheless, it is clear and remarkable that mutation with a single type of mutation operator did not degrade the performance of fuzzers and, on the contrary, gave a high performance improvement on average.
These insights represent a future direction for further research on the effects of mixing different types of mutation operators.

\begin{table*}[tb]
\centering
\caption{Median edge coverage achieved by 6 fuzzers in 24 h with ranks and scores averaged over 10 PUTs. The `*' sign to the right of a value indicates that the sets of the edge coverage obtained by AFL++ and the fuzzer in the column are significantly different in Mann-Whitney's U test ($p < 0.01$).}

\begin{tabular}{lcccccccc}
\toprule

PUT & AFL++ \cite{AFLpp} & MOpt \cite{MOpt} & CMFuzz \cite{CMFuzz} & Karamcheti \cite{AdaptiveFuzzTS} & \HavocMAB{} \cite{HavocMAB} & \OurMethodName{} \\

\midrule

openssl\_x509 & \textit{49.0} & 55.0* & 53.0 & 56.0* & 54.5* & \textbf{57.0}* \\
re2-2014-12-09 & 2280.5 & \textit{2272.0}* & 2278.0 & 2278.5 & \textbf{2281.5} & \textbf{2281.5} \\
proj4-2017-08-14 & 2076.0 & \textit{1956.0}* & 2115.5* & \textbf{2190.5}* & 2075.5 & 2145.5* \\
sqlite3\_ossfuzz & \textit{9201.5} & 9238.0 & 10236.5* & 9926.0* & 9440.0 & \textbf{10819.5}* \\
libxml2-v2.9.2 & 4574.0 & \textit{4233.5}* & 4674.5* & \textbf{4741.0}* & 4702.5* & 4688.5* \\
freetype2-2017 & 5292.0 & \textit{4975.5}* & 5185.0 & 5320.0 & 5298.5 & \textbf{5526.5}* \\
libpcap\_fuzz\_both & \textit{1167.5} & 1213.0 & 1365.0* & 1378.5* & \textbf{1408.0}* & 1351.5* \\
libpng-1.2.56 & 377.5 & 377.0 & \textit{376.0} & 377.0 & 377.5 & \textbf{379.0} \\
lcms-2017-03-21 & 675.0 & 660.5 & 701.0 & 714.0* & \textit{651.0} & \textbf{768.0}* \\
vorbis-2017-12-11 & 770.0 & 769.0 & 770.5 & \textit{767.5}* & 771.0 & \textbf{772.0} \\

\midrule

Rank Avg & 4.4 & \textit{5.1} & 3.9 & 2.7 & 3.0 & \textbf{1.6} \\
Score Avg & 92.81 & \textit{92.13} & 96.37 & 97.59 & 95.69 & \textbf{99.28} \\

\bottomrule

\end{tabular}

\label{tab:vs_existing}
\end{table*}

  \subsubsection{Case Study}
  \label{subsubsec:case-study}

  To examine how SLOPT showed such performance differences, we briefly analyzed two particular PUTs: \texttt{s\-q\-l\-i\-t\-e\-3\-\_\-o\-s\-s\-f\-u\-z\-z} and \texttt{libxml2-v2.9.2}.
  We selected these PUTs because the selection of parameters by SLOPT in them exhibited characteristic trends compared to other methods.
  Table~\ref{tab:sqlite_batch} and \ref{tab:sqlite_mut} show the average percentages of batch sizes and mutation operators, respectively, selected by the bandit optimization methods.
  Because \HavocMAB{} only selects one of two categories (called \textit{unit} and \textit{chunk} mutators) in selecting mutation operators \cite{HavocMAB}, we only display how often each category was selected also for the other methods. 
  
  \noindent\textbf{sqlite3\_ossfuzz}\ \ In this PUT, \OurMethodName{}-AFL++ frequently chose different batch sizes for different seed size as found in Table~\ref{tab:sqlite_batch}a. 
  Although \HavocMAB{}-AFL++ as well as \OurMethodName{}-AFL++ preferred $2$ as batch size the most in total, \OurMethodName{}-AFL++ repeatedly picked $2^6$ for large seeds, which indicates that \OurMethodName{} recognized that larger batch size would make a fuzzing campaign more efficient when mutating large seeds, which is consistent with the observation in Section~\ref{sec:newtarget}.
  In the selection of mutation operators, \OurMethodName{}- and \HavocMAB{}-AFL++ used chunk mutators more often than Karamcheti-AFL++ and CMFuzz-AFL++.
  Considering the fact that \HavocMAB-AFL++ produced lower average code coverage than Karamcheti-AFL++ and CMFuzz-AFL++ in this PUT, the excessive use of chunk mutators possibly reduces the performance improvement when mixing mutation operators. Although \OurMethodName{}-AFL++ also selected chunk mutators excessively, this is not the case with it since it applies exactly one type of mutation operator at once.

  \noindent\textbf{libxml2-v2.9.2}\ \ In this PUT, the four methods all selected each category of mutation operators at similar rates, in comparison to \texttt{sqlite3\_ossfuzz}. However, CMFuzz and \OurMethodName{} were inferior to the other two methods in average code coverage. This would indicate that the difference of even a few percent affects the performance improvement, especially considering that the only difference between CMFuzz- and Karamcheti-AFL++ is the way of selecting mutation operators and that they mutate seeds in completely the same way after the selection of operators.
  Moreover, \OurMethodName- and \HavocMAB{}-AFL++ made a completely different choice of batch sizes. Possible explanations for this would be that \HavocMAB{} blends different types of mutation operators and easily makes generated inputs ill-formed when using large batch sizes unlike \OurMethodName{}, or conversely that \HavocMAB{} can find a new execution path with small batch sizes thanks to blending mutation operators. 
  Note that the seeming inconsistency of this PUT between Figure~\ref{fig:reptition-observations} and Table~\ref{tab:sqlite_batch} comes from the difference between Algorithm~\ref{alg:conv} and \ref{alg:ours}.
  
  \begin{table}[tb]
  \centering
  \caption{Percentages of batch sizes selected by \OurMethodName{} and \HavocMAB{} (a) for \texttt{sqlite3\_ossfuzz} and (b) for \texttt{libxml2-v2.9.2}. The unit is \%.}
  \begin{tabular}{lrrrrrrr}
  \toprule
    (a) sqlite3 & \( 1 \)   & \( 2 \) & \( 2^2 \) & \( 2^3 \) & \( 2^4 \) & \( 2^5 \) & \( 2^6 \) \\
  \midrule
 
\OurMethodName{} $[0,    10^2)$   & \textbf{ 39 } & 22 & 14 & 9 & 6 & 5 & 5 \\
\OurMethodName{} $[10^2, 10^3)$   & 15 & \textbf{ 24 } & 24 & 16 & 10 & 6 & 5 \\
\OurMethodName{} $[10^3, 10^4)$   & 10 & \textbf{ 25 } & 23 & 18 & 11 & 7 & 5 \\
\OurMethodName{} $[10^4, 10^5)$   & 5 & 5 & 6 & 7 & 10 & 21 & \textbf{ 46 } \\
\OurMethodName{} $[10^5, \infty)$ & 9 & 9 & 9 & 11 & 14 & 16 & \textbf{ 31 } \\
\midrule
\OurMethodName{} Overall          & 14 & \textbf{ 23 } & 21 & 16 & 10 & 8 & 9 \\
\HavocMAB{} \cite{HavocMAB}       & 17 & \textbf{ 34 } & 28 & 12 & 5 & 2 & 2 \\

  \bottomrule \\[-0.5em]
 
   \toprule
    (b) libxml2 & \( 1 \)   & \( 2 \) & \( 2^2 \) & \( 2^3 \) & \( 2^4 \) & \( 2^5 \) & \( 2^6 \) \\
  \midrule
 
\OurMethodName{} $[0,    10^2)$   & 7 & 12 & 11 & 11 & 11 & 10 & \textbf{ 38 } \\
\OurMethodName{} $[10^2, 10^3)$   & 3 & 8 & 14 & 16 & 12 & 11 & \textbf{ 36 } \\
\OurMethodName{} $[10^3, 10^4)$   & 3 & 3 & 4 & 5 & 7 & 8 & \textbf{ 70 } \\
\OurMethodName{} $[10^4, 10^5)$   & 5 & 6 & 6 & 7 & 8 & 12 & \textbf{ 56 } \\
\OurMethodName{} $[10^5, \infty)$ & 10 & 10 & 10 & 11 & 11 & 16 & \textbf{ 32 } \\
\midrule
\OurMethodName{} Overall          & 4 & 7 & 10 & 11 & 10 & 10 & \textbf{ 48 } \\
\HavocMAB{} \cite{HavocMAB}       & 21 & \textbf{ 31 } & 26 & 13 & 6 & 2 & 1 \\
  
  \bottomrule
  \end{tabular}
  \label{tab:sqlite_batch}
  \label{tab:libxml_batch}
  
  \end{table}

  \begin{table}[tb]
    \caption{Percentages of mutation operators selected by each fuzzer (a) for \texttt{sqlite3\_ossfuzz} and (b) for \texttt{libxml2-v2.9.2}. The unit is \%.}
    \centering
    \begin{tabular}{lrrrr}
    \toprule
     & \multicolumn{2}{c}{(a) sqlite3} & \multicolumn{2}{c}{(b) libxml2} \\
         & Unit & Chunk & Unit  & Chunk \\

      \midrule
      \OurMethodName{} & 24 & \textbf{76} & 6 & \textbf{94}\\
      \HavocMAB{} \cite{HavocMAB} & 23 & \textbf{77} & 8 &  \textbf{92}\\
      Karamcheti \cite{AdaptiveFuzzTS} & 39 & \textbf{61} & 9 & \textbf{91} \\
      CMFuzz \cite{CMFuzz} & 33 & \textbf{67} & 5 & \textbf{95} \\ 
      \bottomrule
      \end{tabular}
      \label{tab:sqlite_mut}
      \label{tab:limxml_mut}
  \end{table}

\subsection{Effects on Bugs Found}
\label{subsec:magma}

Additionally, we conducted an extensive evaluation with MAGMA to check whether these methods increase a number of bugs found.

\begin{table*}[tb]
  \centering
  \caption{Number of unique bugs found averaged over 10 instances of each fuzzer in 24 h with ranks and scores averaged over 9 targets. The numbers in the parenthesis are the corresponding standard deviations. \#Unique represents the number of unique bugs found by at least 1 instance of each fuzzer.}
  \begin{tabular}{lrrrrrr}
   \toprule

   PUT & 	AFL++ \cite{AFLpp} & MOpt \cite{MOpt} & CMFuzz \cite{CMFuzz}  & Karamcheti \cite{AdaptiveFuzzTS} & \HavocMAB{} \cite{HavocMAB}  & \OurMethodName{}\\
\midrule
openssl & 4.10 (0.54) & 4.00 (0.45) & \textbf{4.30} (0.46) & 4.00 (0.00) & 4.10 (0.83) & 4.20 (0.40)\\
php & 2.40 (0.49) & 2.50 (0.50) & 2.40 (0.49) & \textbf{2.60} (0.49) & 2.40 (0.49) & 2.50 (0.67)\\
sqlite3 & 3.80 (0.60) & 2.80 (0.60) & 4.00 (0.77) & \textbf{4.50} (0.50) & \textbf{4.50} (0.50) & \textbf{4.50} (0.50)\\
libtiff & 8.90 (0.83) & 8.80 (0.98) & 9.00 (1.18) & 9.70 (0.78) & 9.00 (0.63) & \textbf{9.90} (1.04)\\
lua & 0.90 (0.30) & 0.70 (0.46) & 0.60 (0.49) & \textbf{1.00} (0.00) & 0.90 (0.30) & \textbf{1.00} (0.00)\\
libxml2 & 6.50 (0.50) & 6.00 (0.00) & 6.50 (0.67) & \textbf{6.70} (0.64) & \textbf{6.70} (0.64) & 6.50 (0.50)\\
poppler & 10.40 (1.69) & 9.60 (0.92) & 10.10 (1.76) & 9.90 (2.12) & 10.40 (0.66) & \textbf{12.20} (1.78)\\
libpng & 2.70 (0.46) & 2.50 (0.50) & 2.50 (0.50) & \textbf{2.90} (0.30) & 2.70 (0.46) & 2.50 (0.50)\\
libsndfile & 6.90 (0.30) & \textbf{7.00} (0.00) & 6.80 (0.40) & 6.90 (0.30) & 6.90 (0.30) & \textbf{7.00} (0.00)\\
\midrule
Rank Avg & 3.33 & 4.56 & 3.89 & 2.22 & 2.44 & \textbf{1.78}\\
Score Avg & 91.77 & 84.97 & 88.36 & 96.75 & 93.94 & \textbf{97.45}\\
\#Unique & 45 & 43 & 45 & 43 & 45 & \textbf{48}\\

\bottomrule
  \end{tabular}
  \label{tab:magma_crash_comp}
  \end{table*}

Table \ref{tab:magma_crash_comp} lists the average number of bugs triggered by the six fuzzers.
Here, three fuzzers, Karamcheti-, \HavocMAB{}-, and SLOPT-AFL++ ranked higher than AFL++ when compared by average ranks and scores, which is consistent with those metrics in Table~\ref{tab:vs_existing} to some extent.
At the same time, however, the difference between the fuzzers was overall ambiguous compared to code coverage.
Also, we could not see any statistical significance except between MOpt-AFL++ and AFL++ in one PUT.
However, we see this result reasonable, considering that these optimization methods all treat the discovery of a new execution path as a reward and aim at increasing code coverage, not bugs found. In fact, it has recently been recognized that code coverage and the number of bugs found do not have a proportional relationship \cite{FuzzReliability, FuzzBenchAFLppGood}.
Nevertheless, it is worth mentioning that \OurMethodName{} did not cause a noticeable deterioration although it creates a new input always with a single mutation operator, and on the contrary, it found more unique bugs than AFL++ in entirety, as shown in \#Unique in Table~\ref{tab:magma_crash_comp}.

\section{Vulnerability Discovery}

In addition to the evaluations in Section~\ref{sec:evaluation}, we ran \OurMethodName-AFL++ on large real-world programs for a longer period of time.
The purpose of this trial is to see if employing \OurMethodName{} is valuable in practice for identifying bugs, because we found in Section~\ref{subsec:magma} that improving code coverage does not immediately lead to a drastic increase in bugs detected.
Therefore, we ran 10 instances of \OurMethodName-AFL++ for 7 days on 12 PUTs (listed in Table~\ref{tab:commit-ids}) that were manually selected from OSS-Fuzz.
The selection criteria for PUTs were that the PUT must be large in code size so that fuzzers are likely to continuously find new code blocks (thus preventing saturation) and that the project of the PUT must be active and contactable in cases where vulnerabilities are found.
The configuration of each PUT, including initial seeds and dictionaries, was unchanged from those used in the infrastructure of OSS-Fuzz, and the execution environment was the same as Section~\ref{subsec:eval-setup}.

Through manual triage, we found that \OurMethodName-AFL++ discovered 17 bugs, as listed in Table~\ref{tab:7d-bug}. Although we do not deny the possibility of AFL++ being able to find these bugs, we must note that AFL++ has already been fuzzing the PUTs for an enormous number of CPU days on the infrastructure of OSS-Fuzz (estimated to be more than hundreds of CPU years) \cite{OSSFuzzPage, OSSFuzzInFuzzCon}, and thus the possibility is extremely low. In fact, three of these bugs were confirmed as unseen vulnerabilities and were assigned CVE IDs. We believe this result supports the practicability of online optimization methods for code coverage in real-world software testing.

\section{Discussion and Future Work}

\subsection{Threats to Validity}

As in any empirical study, the greatest threats to our claims lie in the correctness and generality of our experimental results, which affect the internal validity and external validity, respectively.

\noindent\textbf{Internal Validity}\ \ The peculiar difficulty in our experiments is that the compared fuzzers share the same understructure.
In Section~\ref{sec:banditcomparison}, the fuzzers are all variations of \OurMethodName-AFL++, with the only difference between them being the bandit algorithm employed.
Similarly, in Section~\ref{sec:evaluation}, the fuzzers are implemented upon AFL++; hence, the seed scheduling or mutation operators themselves are identical.
As we showed in these sections, these fuzzers wholly exhibited performance improvement to some extent, and consequently, it was sometimes difficult to observe evident differences in the quality of their improvement.
To avoid missing these small differences to the extent feasible, we tried to analyze the results of the experiments comprehensively or statistically (as recommended in \cite{FuzzReliability}) while increasing the number of instances of fuzzers and PUTs, within time and resource constraints.
At the same time, we were aware that excessive use of statistical tests can cause multiple testing problems \cite{MultipleComparison, MultipleComparisonVis}, which may lead to spurious conclusions. To avoid these, while finding significant differences as much as possible, we cautiously selected the differences to apply statistical tests to and corrected the resultant p-values using the Holm-Bonferroni method \cite{Holm}.
Another threat to internal validity is the wrong implementation of fuzzers. While implementing them, we referred to their original source code if it was available.
Moreover, for each algorithm, we set up an implementer and a reviewer to ensure that the original algorithm was reproduced. 

\noindent\textbf{External Validity}\ \ To mitigate the threat of the tested PUTs being biased, we ensured the use of various types of real-world PUTs in the evaluations in Section~\ref{sec:evaluation}. For this purpose, we selected PUTs from well-known benchmarks FuzzBench and MAGMA. While we did not adopt all the PUTs in FuzzBench to keep the experimental cost reasonable, we randomly selected 10 PUTs, which accept a wide variety of input file formats as a result.
Another concern regarding external validity is the baseline of the fuzzers. Through all the experiments, we consistently set AFL++ as the suitable baseline because the mutation of AFL++ now consists solely of \textit{havoc} by default \cite{AFLppChangelog}, and AFL++ has high performance among such fuzzers on average according to past benchmarks \cite{FuzzbenchReport}, which is very important for obtaining more rewards and distinguishing optimization methods.
It is possible that the selection of different baselines leads to different conclusions. However, we believe similar results will appear as long as the mutation scheme of a fuzzer is the same because the random mutation \textit{havoc} dominates the performance of the fuzzer, as studied by Wu et al. \cite{HavocMAB}.

\subsection{Correlation of Arms}

One factor that we have not considered when viewing fuzzing as a bandit problem is the correlation between the arms.
For example, in a PUT that accepts only an input consisting of ASCII printable characters, mutation operators that tend to produce non-ASCII printable characters are not considered good choices.
Alternatively, in a PUT that has a limit on the length of inputs to be accepted, mutations such as inserting a constant string or copying a partial byte sequence from another seed are not promising arms. 

While the distributions of rewards are assumed to be independent in standard stochastic bandit problems, there are studies on such problem settings where the arms are correlated, aimed at reducing regret further compared to the standard settings \cite{BanditCorrArm, BanditDepArm, NewAppCorrMAB}.
Even assuming independence, bandit algorithms can greatly improve the efficiency of the fuzzer.
Nonetheless, they suggest the possibility of further improvement by considering this correlation.

\section{Conclusion}
In this study, we explored a new methodology to maximize the effectiveness of bandit algorithms for optimizing the mutation of fuzzers.
We first made the three observations: (a) blending different types of mutation operators into a single mutation is not imperative; (b) batch size is suitable as an optimization target and can have different optimal values depending on seed size; (c) various bandit algorithms give different levels of performance improvement.

Based on these findings, we implemented \OurMethodName-AFL++, which successfully achieved better code coverage than AFL++ in the evaluation.
Also, the performance improvement delivered by \OurMethodName{} was competitive with existing methods.
Although, on the other hand, it did not significantly increase bugs found in the evaluation, \OurMethodName-AFL++ found three previously unseen vulnerabilities in real-world programs from OSS-Fuzz to prove its value in bug identification.

We believe that our results not only provide a general improvement in the performance of existing unoptimized mutation-based fuzzers, but also provide insights to enhance the performance of other fuzzers employing bandit algorithms.
In addition, because our results include practical evaluations conducted with PUTs in OSS-Fuzz, we believe that \OurMethodName-AFL++ can be introduced in OSS-Fuzz with little effort.
We would like to contribute to the introduction of SLOPT-AFL++ in large-scale fuzzing efforts such as Google's OSS-Fuzz initiative.



\begin{acks}
This work was supported by the Acquisition, Technology \& Logistics Agency (ATLA) under the Innovative Science and Technology Initiative for Security 2021 (JPJ004596), and in part by JST ACT-X (JPMJAX200R) and JSPS KAKENHI (JP21J20353).
\end{acks}


\bibliographystyle{ACM-Reference-Format}
\bibliography{reference}

\clearpage
\appendix

\section{Supplemental Tables}


\newcommand\topmidheader[2]{\multicolumn{#1}{c}{\textbf{#2}}\\%
                \addlinespace[1ex]}

\newcommand{\midheader}[2]{%
        \midrule\topmidheader{#1}{#2}}

\newcommand{\specialcell}[3][c]{%
        \begin{tabular}[#1]{@{}#2@{}}#3\end{tabular}}%

\aptLtoX[graphic=no,type=env]{\begin{table}[htb]
  \centering
  \caption{Hyperparameters of bandit algorithms}
  \label{tab:hyperparams}
  \begin{tabular}{llc}
    \toprule
    Sign & Description & Value \\
    \multicolumn{3}{c}{\textbf{UCB1}}\\
    $c$ & Parameter to control the confidence level used in $\sqrt{c \cdot {\log{t}}/{N_t(arm)}}$ & 0.5  \\
    \multicolumn{3}{c}{\textbf{Thompson Sampling}}\\
    $p(\theta)$ & Prior Distribution & $\mathcal{B}(1, 1)$ \\
    \multicolumn{3}{c}{\textbf{discounted Thompson Sampling}}\\
    $\gamma$ & Discount factor & $1-10^{-8}$ \\
    \multicolumn{3}{c}{\textbf{discounted Thompson Samplingadaptive shrinking Thompson Sampling}}\\
    $M$ & Parameter to control memory usage in a data structure ADWIN2 \cite{ADWIN} & 10 \\
    $\delta$ & Parameter to control the confidence level in a data structure ADWIN2 & $1-10^{-7}$ \\
    \multicolumn{3}{c}{\textbf{EXP-IX}}\\
    $\eta_t$ & Parameter used for weights of arms & $\sqrt{\frac{2 \cdot \log{K}}{K \cdot t}}$ \\
    \addlinespace[1ex]
    $\gamma_t$ & Parameter used for loss estimates & $\frac{\eta_t}{2}$ \\
    \multicolumn{3}{c}{\textbf{EXP3++}}\\
    $\alpha$ & Constant used in calculating $\xi_t(a)$ & $3$ \\
    $\beta$ & Constant used in calculating $\xi_t(a)$ & $256$ \\
    \bottomrule
  \end{tabular}
\end{table}}{\begin{table}[htb]
  \centering
  \caption{Hyperparameters of bandit algorithms}
  \label{tab:hyperparams}
  \begin{tabular}{llc}
    \toprule
    Sign & Description & Value \\
    \midheader{3}{UCB1}
    $c$ & \specialcell{l}{Parameter to control the confidence \\ level used in $\sqrt{c \cdot {\log{t}}/{N_t(arm)}}$} & 0.5  \\
    \midheader{3}{Thompson Sampling}
    $p(\theta)$ & Prior Distribution & $\mathcal{B}(1, 1)$ \\
    \midheader{3}{discounted Thompson Sampling}
    $\gamma$ & Discount factor & $1-10^{-8}$ \\
    \midheader{3}{adaptive shrinking Thompson Sampling}
    $M$ & \specialcell{l}{Parameter to control memory usage \\ in a data structure ADWIN2 \cite{ADWIN}} & 10 \\
    $\delta$ & \specialcell{l}{ Parameter to control the confidence \\ level in a data structure ADWIN2} & $1-10^{-7}$ \\
    \midheader{3}{EXP-IX}
    $\eta_t$ & Parameter used for weights of arms & $\sqrt{\frac{2 \cdot \log{K}}{K \cdot t}}$ \\
    \addlinespace[1ex]
    $\gamma_t$ & Parameter used for loss estimates & $\frac{\eta_t}{2}$ \\
    \midheader{3}{EXP3++}
    $\alpha$ & Constant used in calculating $\xi_t(a)$ & $3$ \\
    $\beta$ & Constant used in calculating $\xi_t(a)$ & $256$ \\
    \bottomrule
  \end{tabular}
\end{table}}

\begin{table}[htb]
  \centering
  \caption{Commit IDs of the PUTs used in our vulnerability discovery and AFL++ used as the baseline.}
  \begin{tabular}{lc}
    \toprule
    Program & Commit \\
    \midrule

    AFL++ & 32a0d6ac315 (ver ++3.14c) \\
    Bloaty &  60209eb \\
    HarfBuzz & 77eeec5 \\
    libarchive & 86c9361 \\
       libxml2 & dea91c9 \\
    MuPDF & ef3d68d \\
   PHP & fdf0455f \\
    Poppler & 6d72d82 \\
    PROJ & 76dfefe \\
    QPDF &  3794f8e \\
    libtpm2 & bc3bb26 \\
    Wireshark  & 1fc621e \\
    Xpdf & N/A (ver 4.03) \\

    \bottomrule
  \end{tabular}
\label{tab:commit-ids}
\end{table}

\begin{table}[htb]
  \centering
  \caption{Initial and theoretical maximum values of code coverage of the PUTs in FuzzBench. 
           Initial values were investigated only in the PUTs used.}
  \begin{tabular}{lcc}
    \toprule
    PUT & Initial & Maximum \\
    \midrule

bloaty\_fuzz\_target & N/A & 83114 \\
curl\_curl\_fuzzer\_http & N/A & 78362 \\
freetype2-2017 & 1517 & 26262 \\
harfbuzz-1.3.2 & N/A & 12212 \\
jsoncpp\_jsoncpp\_fuzzer & N/A & 2114 \\
lcms-2017-03-21 & 149 & 7036 \\
libjpeg-turbo-07-2017 & N/A & 9384 \\
libpcap\_fuzz\_both & 2 & 7294 \\
libpng-1.2.56 & 138 & 3736 \\
libxml2-v2.9.2 & 258 & 67994 \\
libxslt\_xpath & N/A & 51456 \\
mbedtls\_fuzz\_dtlsclient & N/A & 12888 \\
openssl\_x509 & 6026 & 54116 \\
openthread-2019-12-23 & N/A & 19846 \\
php\_php-fuzz-parser & N/A & 215210 \\
proj4-2017-08-14 & 46 & 6534 \\
re2-2014-12-09 & 1 & 3982 \\
sqlite3\_ossfuzz & 4767 & 28766 \\
systemd\_fuzz-link-parser & N/A & 1798 \\
vorbis-2017-12-11 & 410 & 4082 \\
woff2-2016-05-06 & N/A & 5708 \\
zlib\_zlib\_uncompress\_fuzzer & N/A & 910 \\

    \bottomrule
  \end{tabular}
\label{tab:fuzzbench_max_cov}
\end{table}

\begin{table}[htb]
\centering
\caption{List of unique bugs found in the 7-day trial (manually triaged).}
\begin{minipage}{\columnwidth}

\centering
\begin{tabular}{lll}
\toprule

ID & PUT & Bug Type \\
\midrule
Bug-A & bloaty & NULL Pointer Deref \\
Bug-B & harfbuzz & Out-of-bounds Read \\
Bug-C & mupdf & Assertion Fail \\
Bug-D & mupdf & NULL pointer deref \\
Bug-E & xpdf & Stack Overflow \\
Bug-F & xpdf & NULL Pointer Deref \\
Bug-G \footnote{CVE-2022-24106 is issued.} & xpdf & Use of Uninitialized Value \\
Bug-H \footnote{CVE-2022-24107 is issued.} & xpdf & Integer Overflow \\
Bug-I & php & Use-After-Free \\
Bug-J & php & Use-After-Free \\
Bug-K & php & NULL Pointer Deref \\
Bug-L & php & Use-After-Free \\ 
Bug-M & php & NULL Pointer Deref \\
Bug-N & php & Assertion Fail \\
Bug-O & php & Use-After-Free \\
Bug-P & php & Use-After-Free \\
Bug-Q \footnote{CVE-2022-23308 is issued.} & libxml2 & Use-After-Free \\
\bottomrule
\end{tabular}

\label{tab:7d-bug}
\end{minipage}
\end{table}

\begin{table*}[htb]
  \centering
  \caption{List of the PUTs used in Section~\ref{sec:banditcomparison}. If the source code of a PUT was maintained in Git, the latest version at the time of the experiment in the master (or main) branch was used for the build. The `+' sign in a version indicates that the used source code is not the official release version of the source code.}
  \renewcommand\tabularxcolumn[1]{m{#1}}
  \renewcommand{\arraystretch}{1.2}
  \begin{tabularx}{\textwidth}{lXllXc}
    \toprule
    Project & Version & Commit ID & PUT & Format of Initial Seeds & Initial Edge Coverage \\
    \midrule
    Bloaty & v1.1+ & 60209eb & fuzz\_target & Executable (e.g., ELF, PE, Mach-O) & 4773\\
    libmpeg2 & N/A & 5432dc1 & mpeg2\_dec\_fuzzer & MPEG2 & 2428 \\
    PHP & 8.0+ & fdf0455f & php-fuzz-execute & PHP source code & 25241 \\
    HarfBuzz & 3.1.0 & 77eeec5 & hb-shape-fuzzer & Font (e.g., TrueType, OpenType) & 15298 \\
    Xpdf & 4.03 & N/A & fuzz\_pdfload & PDF & 4755 \\
    libtpm2 & N/A & bc3bb26 & tpm2\_execute\_command\_fuzzer & TPM command & 3884\\
    libyaml & v0.2.5+ & f8f760f & libyaml\_dumper\_fuzzer & YAML & 1310 \\
    libzip & 1.8.0+ & bff2eb9 & zip\_read\_fuzzer & ZIP & 805 \\
    libgit2 & v1.3.0+ & 50b4d53 & download\_refs\_fuzzer & Git packet & 3911 \\
    file & 5.41+ & fcbb5d8 & magic\_fuzzer & any (e.g., Zstd compressed file) & 1171 \\
    \bottomrule
  \end{tabularx}
\label{tab:put_details}
\end{table*}



\begin{table*}[htb]
\centering
\caption{Median edge coverage obtained by AFL++ and 8 versions of \OurMethodName-AFL++ in 10 PUTs after 24 h. }

\begin{tabular}{lccccccccc}
\toprule

PUT & AFL++ & UCB1 & KLUCB & TS & dTS & dBE & ADS-TS & EXP3-IX & EXP3++ \\
\midrule

bloaty & \textit{1845.5} & 2198.5 & 2246.0 & 2232.5 & 2191.0 & 2292.0 & \textbf{2340.0} & 2181.5 & 2231.5 \\
harfbuzz & \textit{13497.5} & 14031.5 & 14247.5 & 14360.5 & \textbf{14374.0} & 14067.5 & 14149.0 & 13883.0 & 13891.0 \\
xpdf & \textit{3384.0} & 3494.0 & 3812.5 & \textbf{4618.5} & 4166.5 & 3791.5 & 3902.0 & 3860.0 & 3615.0 \\
libzip & \textit{267.5} & 272.0 & 274.0 & 268.0 & 268.5 & 271.5 & \textbf{276.0} & 271.5 & 268.0 \\
libgit2 & 898.0 & 888.5 & 890.5 & 906.5 & \textbf{916.0} & 884.0 & 914.0 & 899.5 & \textit{881.0} \\
php & \textit{9841.5} & 11861.0 & 13551.5 & \textbf{14324.0} & 14187.5 & 12657.5 & 13408.0 & 11423.5 & 11828.5 \\
libmpeg2 & \textit{1873.5} & 1900.5 & 1905.0 & 1905.5 & \textbf{1906.5} & 1903.0 & \textbf{1906.5} & 1897.0 & 1902.0 \\
tpm2 & \textit{281.5} & 299.5 & 313.0 & 317.0 & \textbf{317.5} & 305.0 & 311.0 & 298.5 & 291.0 \\
libyaml & 2811.5 & 2841.0 & \textbf{2841.5} & \textit{2800.5} & 2837.0 & 2827.5 & 2831.5 & 2828.0 & 2834.5 \\
file & 830.5 & 829.5 & 828.0 & 827.0 & 827.5 & 833.5 & \textbf{840.5} & 826.5 & \textit{826.0} \\

\bottomrule

\end{tabular}

\label{tab:alg_cmp_all}
\end{table*}

\begin{table*}[htb]
\centering
\caption{P-value of Mann-Whitney's U test (Holm-Bonferroni corrected) and Vargha-Delaney's $\hat{A}_{12}$ between AFL++ and the fuzzer in the column for the evaluation conducted in Section~\ref{subsec:eval-vs-existing}. If the p-value is bold, the difference is significant in the test ($p < 0.01$). The characters `L', `M', `S' and `N' in parentheses indicate that the effect size is large, medium, small, and none, respectively, according to \cite{A12}. The `+' sign means the fuzzer in the column is superior to AFL++ when compared by rank sum as well as $\hat{A}_{12}$, and the `-' sign means the opposite.}
\begin{tabular}{lllllllllllll}
 \toprule

  & \multicolumn{2}{c}{MOpt} & \multicolumn{2}{c}{CMFuzz} & \multicolumn{2}{c}{Karamcheti} & \multicolumn{2}{c}{\HavocMAB{}} & \multicolumn{2}{c}{SLOPT} \\
  \cmidrule(r){2-3}\cmidrule(r){4-5}\cmidrule(r){6-7} \cmidrule(r){8-9} \cmidrule(r){10-11}
  PUT & $p$ & $\hat{A}_{12}$ & $p$ & $\hat{A}_{12}$ & $p$ & $\hat{A}_{12}$ & $p$ & $\hat{A}_{12}$ & $p$ & $\hat{A}_{12}$ \\
\midrule

openssl\_x509 & \textbf{ < 0.001 } & 0.82 (+L) & \textbf{ 0.023 } & 0.71 (+L) & \textbf{ < 0.001 } & 0.92 (+L) & \textbf{ < 0.001 } & 0.82 (+L) & \textbf{ < 0.001 } & 0.91 (+L) \\
re2-2014-12-09 & \textbf{ < 0.001 } & 0.18 (-L) & > 0.1 & 0.37 (-S) & > 0.1 & 0.38 (-S) & > 0.1 & 0.47 (-N) & > 0.1 & 0.52 (+N) \\
proj4-2017-08-14 & \textbf{ < 0.001 } & 0.08 (-L) & \textbf{ < 0.001 } & 0.86 (+L) & \textbf{ < 0.001 } & 0.99 (+L) & > 0.1 & 0.54 (+N) & \textbf{ < 0.001 } & 0.92 (+L) \\
sqlite3\_ossfuzz & > 0.1 & 0.55 (+N) & \textbf{ < 0.001 } & 0.85 (+L) & \textbf{ < 0.001 } & 0.93 (+L) & 0.1 & 0.68 (+M) & \textbf{ < 0.001 } & 1.00 (+L) \\
libxml2-v2.9.2 & \textbf{ < 0.001 } & 0.08 (-L) & \textbf{ < 0.001 } & 0.93 (+L) & \textbf{ < 0.001 } & 0.98 (+L) & \textbf{ < 0.001 } & 0.97 (+L) & \textbf{ < 0.001 } & 0.84 (+L) \\
freetype2-2017 & \textbf{ < 0.001 } & 0.08 (-L) & 0.094 & 0.33 (-M) & > 0.1 & 0.54 (+N) & > 0.1 & 0.52 (+N) & \textbf{ < 0.001 } & 0.79 (+L) \\
libpcap\_fuzz\_both & > 0.1 & 0.57 (+S) & \textbf{ < 0.001 } & 0.79 (+L) & \textbf{ < 0.001 } & 0.80 (+L) & \textbf{ < 0.001 } & 0.87 (+L) & \textbf{ < 0.001 } & 0.81 (+L) \\
libpng-1.2.56 & > 0.1 & 0.42 (-S) & > 0.1 & 0.36 (-M) & > 0.1 & 0.49 (-N) & > 0.1 & 0.56 (+S) & 0.049 & 0.68 (+M) \\
lcms-2017-03-21 & > 0.1 & 0.45 (-N) & \textbf{ 0.037 } & 0.70 (+M) & \textbf{ < 0.001 } & 0.85 (+L) & > 0.1 & 0.37 (-S) & \textbf{ < 0.001 } & 0.88 (+L) \\
vorbis-2017-12-11 & > 0.1 & 0.39 (-S) & > 0.1 & 0.56 (+S) & \textbf{ < 0.001 } & 0.20 (-L) & > 0.1 & 0.62 (+S) & 0.092 & 0.65 (+M) \\

\bottomrule
\end{tabular}
\label{tab:statistics}
\end{table*}

\clearpage

\section{Algorithm Overview}

\begin{algorithm}[H]

\centering
\caption{Pseudocode of \OurMethodName{}}
\label{alg:slopt}

\begin{algorithmic}[0]

\Require{\mbox{}\\
    $initial\_seeds$ -- a set of initial test cases \\
    $program$ -- a PUT to be fuzzed
}

\Ensure{\mbox{}\\
    $queue$ -- a set of valuable test cases \\
    $crashes$ -- a set of test cases that trigger crashes
}

\vspace{5pt}

\Function{RandomMutation}{$seed, instance_{mut}, instances_{bat}$}
\State $input$ $\gets$ \Call{CopyBytesFromSeed}{$seed$}
\State $mutation$ $\gets$ \Call{SelectArm}{$instance_{mut}$}
\State $idx$ $\gets$ \Call{GetGroupIndex}{$len(input)$}
\State $batch\_size$ $\gets$ \Call{SelectArm}{$instances_{bat}[idx][mutation]$}
\For{$i$ $\gets$ $1$ \textbf{to} $batch\_size$}
    \State $pos$ $\gets$ \Call{SelectPosition}{$input$}
    \State $input$ $\gets$ \Call{ApplyOperator}{$mutation, input, pos$}
\EndFor
\State \textbf{return} $input, mutation, batch\_size$
\EndFunction




\vspace{5pt}

\Function{MutationFuzzing}{$initial\_seeds, program$}

\State $crashes$ $\gets$ $\varnothing$
\State $queue$ $\gets$ \Call{ConstructQueue}{$initial\_seeds$}
\State $instance_{mut}$ $\gets$ \Call{CreateBanditArms}{$number\_of\_mutations$}
\For{$i$ $\gets$ $1$ \textbf{to} $5$}
 \For{$j$ $\gets$ $1$ \textbf{to} $number\_of\_mutations$}
  \State $instances_{bat}[i][j]$ $\gets$ \Call{CreateBanditInstance}{$7$}
 \EndFor
\EndFor

\State

\While{ $\neg$ \Call{UserWantsStop}{\null}}
 \State $seed$ $\gets$ \Call{SelectSeed}{$queue$}
 \State $energy$ $\gets$ \Call{DecideEnergy}{$seed$}
 \For{$i$ $\gets$ $1$ \textbf{to} $energy$}
  \State $input, mutation, batch\_size$ 
  \State $\gets$ \Call{RandomMutation}{$seed, instance_{mut}, instances_{bat}$}
  \State $result$ $\gets$ \Call{ExecutePUT}{$program, input$}
  \State $b$ $\gets$ \Call{WasInputValuable}{$result$}
  \State \Call{RewardArm}{$mutation, b$}
  \State \Call{RewardArm}{$batch\_size, b$}
  \State \Call{SaveInputIfValuable}{$queue, input, result$}
  \State \Call{SaveInputIfCrash}{$crashes, input, result$}
 \EndFor
\EndWhile
\EndFunction


\end{algorithmic}
\end{algorithm}


\end{document}